\documentclass[fleqn,12pt]{wlscirep}
 
\usepackage[utf8]{inputenc}
\usepackage[T1]{fontenc}
\usepackage{setspace}
\usepackage{siunitx}
\title{Broadband on-chip single-photon spectrometer}

\author[1]{Risheng Cheng}
\author[1,2]{Chang-Ling Zou}
\author[1]{Xiang Guo}
\author[1]{Sihao Wang}
\author[1]{Xu Han}
\author[1,*]{Hong X. Tang}

\affil[1]{Department of Electrical Engineering, Yale University, New Haven, Connecticut 06511, USA.}
\affil[2]{Department of Optics and Optical Engineering, University of Science and Technology of China, Hefei 230026, Anhui, China}
\affil[*]{hong.tang@yale.edu}
\begin{abstract}
\end{abstract}

\begin{document}
\flushbottom
\maketitle
\thispagestyle{empty}

\noindent\textbf{Single-photon counters are single-pixel binary devices that click upon the absorption of a photon but obscure its spectral information, whereas resolving the colour of detected photons has been in critical demand for frontier astronomical observation, spectroscopic imaging and wavelength division multiplexed quantum communications. Current implementations of single-photon spectrometers either consist of bulky wavelength-scanning components or have limited detection channels, preventing parallel detection of broadband single photons with high spectral resolutions. Here, we present the first broadband chip-scale single-photon spectrometer covering both visible and infrared wavebands spanning from $\mathbf{600}\,\mathbf{nm}$ to $\mathbf{2000}\,\mathbf{nm}$. The spectrometer integrates an on-chip dispersive echelle grating with a single-element propagating superconducting nanowire detector of ultraslow-velocity for mapping the dispersed photons with high spatial resolutions. The demonstrated on-chip single-photon spectrometer features small device footprint, high robustness with no moving parts and meanwhile offers more than 200 equivalent wavelength detection channels with further scalability.
}
\vspace{0.2cm}

\noindent High-performance single-photon spectrometers are among the most sought-after instruments in cutting-edge research fields especially for applications in photon-scarce environments. For example, in the applications such as astronomical spectroscopy\cite{appenzeller2012introduction}, fluorescence imaging \cite{lichtman_2005_fluorescence_review,camp_2015_Raman_review} and remote sensing, the signal light is extremely faint, and thus single-photon sensitive spectrometers with low dark count noise are crucial. In wavelength division multiplexed (WDM) quantum communications\cite{ciurana_2014_quantum_WDM,dynes_2016_WDM_quantum_information,wengerowsky_2018_WDM_quantum_communication, eriksson_2019_wdm_qkd}, advanced single-photon detectors with  spectral resolvability combining low dark counts, fast speed, and high timing resolution are ideal devices as the quantum receivers. However, current implementations of single-photon spectrometers consist of bulky wavelength-scanning components and photomultiplier tubes (PMT) or semiconductor-based single-photon counters of finite channels \cite{gudkov_2015_PMT_array,finocchiaro_2007_spad_spectrometer,gudkov_2013_fluorescent_detection_spectromter}, hindering parallel detection of broadband photon input with high spectral resolutions. Moreover, the semiconductor detectors, such as InGaAs single-photon avalanche diodes (SPADs) used in telecom-band photon counting, also suffer from huge dark counts, limited efficiency, slow speed and after-pulsing\cite{Hadfield_2009_SPD_review}. 

On the other hand, superconducting nanowire single-photon detectors (SNSPDs)\cite{gol_2001_first_SNSPD,natarajan_2012_SNSPD_review} have recently emerged as one of the best alternatives, outperforming the semiconductor counterparts in all aspects with near-unity efficiency, high speed, low jitter, low dark counts \cite{marsili_2013_93p_efficiency,esmaeil_2017_92p_nbn_detector_Delft,zhang_2017_92p_nbn_detector,korzh_2018_low_jitter_JPL,schuck_2013_mHz_dark_count} and the capability of on-chip integration with integrated nanophotonic circuits\cite{Fiore_2011_waveguide_snspd,pernice_2012_waveguide_SNSPD,MIT_2015_on_chip_detector,akhlaghi_2015_perfect_nanowire_absorber,schuck_2016_on_chip_HOM,wolfram_2016_fully,Bristol_2016_modelling_snspd_wg_cavity,wolfram_2018_2DPC,ferrari_2018_waveguide_snspd_review,zhu_2018_scalable_detector_MIT}. However, these detectors operate in a strong non-linear mode -- only informing the presence or absence of photons -- and thus cannot discriminate the energy or provide the spectral information of the detected photons. Waveguide-based structure is proposed to circumvent this problem\cite{kahl_2017_spectrally_multiplexed_snspd}, where up to eight parallel SNSPDs are co-integrated with an arrayed waveguide grating (AWG) and used for the fluorescence imaging of colour centres in diamond. Yet, individual readout scheme is employed here for the discrete detector array, which ultimately limits the scalability of the detection channels as well as further improvement of the spectral resolution and operation bandwidth. In another research on fibre-assisted spectrometers \cite{toussaint_2015_Raman_SNSPD,gerrits_2015_spectral_correlation_NIST}, SNSPDs are employed in conjunction with very long fibres to convert the arrival time of detected photons to the wavelength information based on the large dispersion introduced by the fibres. However, this scheme only works with pulsed photon sources, and the complete set-up comprises discrete components that remain to be integrated.  

Here, we propose and experimentally implement a broadband on-chip single-photon spectrometer that overcomes all the above-mentioned challenges. By interfacing a millimetre-size nanophotonic echelle grating with a single-element meandered SNSPD, we realize continuous mapping of the spectral information of dispersed input photons (Fig.$\,$1a). The meandered SNSPD is capped with a high-\textit{k} (high dielectric constant) layer to form a microstrip transmission line with a group velocity as low as 0.0073\textit{c} (\textit{c}: the speed of light in vacuum) for precision time-tagging, which has been recently employed to realize a single-photon imager \cite{zhao_2017_single_photon_imager}. With a continuous NbN nanowire of a total length of $7\,\mathrm{mm}$, we demonstrate more than 200 effective single-photon detection channels over a broad wavelength range between $600\,\mathrm{nm}$ and $2000\,\mathrm{nm}$. This on-chip spectrometer uniquely combines the benefits of planar nanophotonic and superconducting nanowire circuits and therefore is inherently scalable. Future scale-up fabrication of our design at wafer dimensions could further enhance the spectral resolution and the number of detection channels while maintaining the same integration and readout architecture. 

\section*{{\normalsize{}Results}}
\textbf{Principle and device design.}
The dispersive component of our device is based on the echelle grating, which is known in astronomical and precision spectroscopy for its high dispersive power. With the advent of integrated photonics, it becomes feasible to lithographically define millimetre-diameter echelle gratings on a chip and effectively disperse the incoming photons from an input waveguide, as recently demonstrated in silicon photonic structures\cite{sciancalepore_2015_SOI_spectrometer}. Here, we implement the echelle spectrometer in stoichiometric silicon nitride (Si$_3$N$_4$) to allow for broadband optical waveguiding from visible to mid-infrared.  In Fig.$\,$1e, we present the schematic of the standard Rowland mounting\cite{lycett_2013_echelle_grating} to illustrate the operation principle. The facets of a flat grating are projected onto a circular section to form a focusing grating with the radius equal to the diameter of the Rowland circle, upon which the input waveguide and the superconducting nanowire are lithographically mounted. The focusing grating and the Rowland circle are internally tangent at the centre of the grating. Photons are first edge-coupled from the lensed fibre into the tapered Si$_3$N$_4$ waveguide (Fig.$\,$1c) and then diverge in the free propagation region of the dielectric slab within the Rowland circle (Fig.$\,$1d). Afterwards, the photons are reflected, diffracted, and refocused by the concave grating to the focusing point \textit{P} on the superconducting nanowire (Fig.$\,$1e), the position of which varies with the wavelength of the input photons. The diffraction angle is determined by the equation
\begin{equation}
\hspace{0.28\textwidth} d(\mathit{\mathrm{sin}}(\theta_\mathrm{in})+\mathrm{sin}(\theta_{m}))=\frac{m\lambda}{n_\mathrm{eff}}
\end{equation}

\noindent where \textit{d} is the period of the grating, \textit{$\theta_\mathrm{in}$} the angle of incidence, \textit{$\theta_{m}$} the angle of the \textit{$m^\mathrm{th}$} order diffraction, \textit{$\lambda$} the wavelength of the incident photons in free space, and\textit{ $n_\mathrm{eff}$} the effective refractive index of the mode in the slab waveguide. The blaze angle of the grating facets can be tuned to guide the diffracted light into the desired order. 

While inheriting all conventional merits of SNSPDs, our nanowire detector features an ultralow-velocity microwave delay line formed by capping the NbN superconducting nanowire with high-\textit{k} dielectric material (alumina or AlO\textit{\textsubscript{x}}) and top metal ground (aluminium or Al) as shown in Fig.$\,$1b. Due to the very large kinetic inductance of the nanowire and the slow microstrip line design, the pair of photon-excited microwave pulses of opposite polarities propagate slowly at a group velocity as low as 0.73\% of the vacuum speed along the circumference of the Rowland circle. It is worth noting that this is the slowest results reported among the superconducting nanowire delay lines and twice slower than oxide-cladded delay lines achieved previously \cite{zhu_2018_scalable_detector_MIT} (see Supplementary Note 3). As a result, the times of arrival to the amplifiers attached at the both ends of the nanowire transmission line could be registered with high temporal resolution. This time-tagged single-photon signal or the arrival time difference \textit{$\Delta t=t_{1}-t_{2}$} can be used to trace out the spatial distribution of incident photons. In our nanophotonic devices, this spatial distribution in turn arises from dispersed photons of different colours by the echelle grating with high spatial resolution (Fig.$\,$1f). Therefore, our design permits a single-element nanowire to function as a multi-channel spectral-resolving single-photon detector.


To demonstrate the proof of principle and modes of operation, multiple devices of varying size and design parameters are fabricated on the same chip and cooled down to $1.5\,\mathrm{K}$ temperature in a dilution refrigerator for characterizations. The edge-coupling scheme between the fibre and the device chip based on our cryogenic active alignment set-up could guarantee broad spectrum input coupling from visible to infrared waveband. We categorize the devices into two main designs: (i) \textit{broadband} design based on a \SI{400}{\micro\meter}-radius Rowland circle and targeting 600-$2000\,\mathrm{nm}$ waveband; (ii) \textit{telecom-band} devices using a larger 1.6$\,$mm-radius Rowland circle and dedicatedly designed for telecommunication waveband between 1420-$1640\,\mathrm{nm}$. More details on the design parameters, device fabrication and characterization can be found in the Methods section and Supplementary Note 1 and 2. The optical micrograph and scanning electron micrograph (SEM) images for one of the broadband devices are shown in Fig.$\,$2a-2d. The detector part consists of a long 60$\,$nm-wide meandering NbN wire, both ends of which are gradually tapered to microns width to form Klopfenstein-type impedance tapers that help preserve the fast-rising edges of photon-excited microwave pulses.    

\noindent \textbf{Broadband device.}
Figure 3a presents 2.5-dimensional finite-difference time-domain (FDTD) simulation results of the broadband device. As expected, the diffracted light could be well refocused on the Rowland circle where the superconducting nanowire is placed. For shorter wavelengths, higher-order diffraction modes (\textit{m}>1) exist, but they are effectively suppressed at least one order of magnitude lower than fundamental modes by tuning the blaze angle of the grating. The $0^\mathrm{th}$ order mode (\textit{m}=0) represents the direct reflection of the input beam by the grating without diffraction, which do not have wavelength discriminating effect and are not desired. The nanowire detector therefore is not mounted in this diffraction region. In Fig.$\,$3b, we plot the normalized histogram of photon counts versus \textit{$\Delta t$} measured for different wavelength input photons (see Supplementary Note 7 for the raw data). All the major peaks are from TE modes, while the minor peaks marked by dashed circles are due to the TM modes, \textit{$n_\mathrm{eff}$} of which are always smaller than that of TE modes. The TM modes are excited at shorter wavelengths due to the multimode operation of the input waveguide, while they are suppressed for longer wavelengths where the waveguide only supports single TE modes. Note that such imperfection could be resolved in future optimizations of the device by introducing an on-chip TE-pass polarizer to the input waveguide. The full width at half maximum (FWHM) of the major peaks are 13-$19\,\mathrm{ps}$, which corresponds to a better than $7\,\mathrm{nm}$ spectral resolution and suggests more than 200 wavelength detection channels between 600-$2000\,\mathrm{nm}$. In Fig.$\,$3c, \textit{$\Delta t$} corresponding to the major peaks are extracted and plotted versus the wavelength, which agrees well with the diffraction angles obtained from the FDTD simulation results (see Fig.$\,$1e for the angle definition). Figure$\,$3d shows the normalized photon counting rates measured as a function of the relative bias current to the switching current ($I_\mathrm{bias}/I_\mathrm{SW}$) at wavelengths from $750\,\mathrm{nm}$ to $1970\,\mathrm{nm}$. Apparent saturation behaviour is observed for all the curves, suggesting a near-unity internal quantum efficiency of the nanowire detector over the whole spectrum\cite{marsili_2013_93p_efficiency,esmaeil_2017_92p_nbn_detector_Delft,zhang_2017_92p_nbn_detector}. To the best of our knowledge, it is the first time to realize such a compact on-chip single-photon spectrometer simultaneously covering visible and infrared wavebands.      

\noindent \textbf{Telecom-band device.}
The spectral resolving power of our on-chip single photon spectrometer can be further assessed with a telecom-band device design, which has four times larger Rowland circle radius and uses $6^\mathrm{th}$ order diffraction for enhanced dispersive effect. Figure$\,$4a shows the normalized histogram of photon counts versus \textit{$\Delta t$}. The FWHM of each peak is 25-$30\,\mathrm{ps}$, corresponding to a resolution of 2.5-$3\,\mathrm{nm}$ in wavelength. In order to further evaluate the spectral resolution, we combine two continuous-wave (CW) laser beams with wavelengths separated by $2.5\,\mathrm{nm}$ through a fibre-splitter and send them to the device after an appropriate attenuation. As shown in Fig.$\,$4b, there are two distinctly resolved peaks, consistent with the projected resolution for single photons. Due to the continuous and sub-wavelength structure of the nanowire detector, the measurement precision of $\Delta t$ can be always boosted by repetitive measurement, which indicates that our spectrometer device can also work in wavelength-meter mode to provide a resolution well beyond the aforementioned value when measuring the wavelength of single-colour input photons. The histogram curves shown in Fig.$\,$4c displays the response from light inputs of $0.1\,\mathrm{nm}$ wavelength difference. We note that the resolution in this operation mode is not determined by the FWHM values of the histogram peaks but limited by the stability of laser source and the long integration time of the histogram measurement associated with the limited acquisition speed and refresh rate of the oscilloscope. By further scaling our device and utilizing application-specific high speed correlation electronics as oppose to an oscilloscope, we expect that the wavelength resolution could be improved to picometre level. We also characterize the timing performance of the spectrometer by recording the histogram of photon counts as a function of the arrival time difference between the detector signal and the synchronization signal of a 2.4$\,$ps-pulsed laser. The results are shown in Fig.$\,$4d with $40\,\mathrm{ps}$ jitter defined as the FWHM of the histogram profile, which is consistent with conventional SNSPDs. More details on the jitter characterization and the impact of noise-introduced timing jitter on spectral resolution are provided in Supplementary Note 6 and 5. 

\section*{{\normalsize{}Discussion}}
The ultimate resolution of our single photon spectrometer is set by the achievable size of the grating which is eventually limited by the wafer size and also the total length of the nanowire delay line to cover all the desired diffraction angles. As detailed in Supplementary Note 8, we estimate that $100\,\mathrm{pm}$ resolution with $200\,\mathrm{nm}$ bandwidth and thus 2,000 wavelength channels is feasible with a telecom-band device design based on a 50$\,$mm-radius Rowland circle and a 40$\,$mm-long nanowire. In order to realize such a device, although we anticipate that the superconducting nanowire circuit could be fabricated without degradation as previously demonstrated in large-area  SNSPDs\cite{allmaras_2017_large_area_wsi_detector}, some technical challenges remain to be solved on the photonics side, such as pattern decoherence induced by thermal expansion and stitching error during the electron-beam lithography of large nanophotonic structure across distant writing fields. Future improvement also requires the optimization of the device design to improve the system detection efficiency (see Supplementary Note 1 and 4) and remove the non-Gaussian tails in the histogram results to increase the dynamic range (see Supplementary Note 7). It is also notable that although this work focuses on the single-photon detection regime, our on-chip spectrometer can also be extended to study two-photon absorption events \cite{zhu_2018_scalable_detector_MIT} and thus could be an important tool for characterizing spectrally  entangled photon pairs. Moreover, our design can be easily extended to mid-infrared waveband, where lots of important and fast-emerging applications reside, such as remote sensing and single-photon lidar. We envision that our spectrometers will find important and immediate applications in quantum sensing, communication and frontier spectroscopic imaging technologies. 


\section*{{\normalsize{}Online content}}

\noindent {\footnotesize{}Any methods, additional reference, Nature Research reporting summaries, source data, statements of data availability and associated accession codes are available in the online version of the paper.}{\footnotesize\par}

\setstretch{1.0}


\setstretch{1.5}

\section*{{\normalsize{}Acknowledgements}}

{\footnotesize{}We acknowledge funding support from DARPA DETECT program through an ARO grant (No: W911NF-16-2-0151), NSF EFRI grant (EFMA-1640959), AFOSR MURI grant (FA95550-15-1-0029), and the Packard Foundation.  The authors would like to thank Michael Power, James Agresta, Christopher Tillinghast, and Dr. Michael Rooks for their assistance provided in the device fabrication. The fabrication of the devices was done at the Yale School of Engineering \& Applied Science (SEAS) Cleanroom and the Yale Institute for Nanoscience and Quantum Engineering (YINQE).}{\footnotesize\par}

\section*{{\normalsize{}Author contributions }}

{\footnotesize{}R.C., C.-L.Z., X.G., and H.X.T. conceived the idea and experiment; R.C. designed the devices; R.C. and S.W. fabricated the device; R.C. and X.H. performed the measurements; R.C., C.-L.Z. and X.G. analysed the data. R.C., C.-L.Z. and H.X.T. wrote the manuscript with input from all authors. H.X.T. supervised the project.}{\footnotesize\par}

\section*{{\normalsize{}Competing interests }}

{\footnotesize{}The authors declare no competing interests. }{\footnotesize\par}

\section*{{\normalsize{}Additional information}}
{\footnotesize{}\textbf{Supplementary information} is available in the online version of the paper.}

{\noindent\footnotesize{}\textbf{Correspondence and requests for materials} should be addressed to H.X.T.}
{\footnotesize\par}

{\footnotesize{}\newpage}{\footnotesize\par}

\section*{{\normalsize{}Methods}}

\textbf{Device fabrication.} The photonic circuit components are patterned from 330$\,\mathrm{nm}$-thick stoichiometric Si$_3$N$_4$ on Si wafers covered with \SI{3.3}{\micro\meter}-thick thermally-grown oxide. The superconducting detectors and microwave circuits are realized in a 8\,nm-thick NbN thin film, which is deposited on the Si$_3$N$_4$ with alignment markers (10\,nm Cr\,/\,100\,nm Au) fabricated in advance. After the NbN film deposition, we define superconducting nanowires along with impedance tapers by the exposure of negative-tone 6\% hydrogen silsesquioxane (HSQ) resist using high-resolution ($100\,\mathrm{kV}$) electron-beam lithography (EBL) and the subsequent development in tetramethylammonium hydroxide (TMAH)-based developer MF-312. In a second EBL step, electrode pads are defined using double-layer polymethyl methacrylate (PMMA) positive-tone resist. After the development in methyl isobutyl ketone (MIBK) and isopropyl alcohol (IPA), we liftoff electron-beam evapoarted 10$\,$nm Cr adhesion layer and 100$\,$nm Au in acetone to form the contact pads. Later, the HSQ nanowire pattern is transferred to the NbN layer in a timed reactive-ion etching (RIE) step employing tetrafluoromethane (CF\textsubscript{4}) chemistry. In a third EBL step, we expose the positive-tone ZEP520A polymer resist for the photonic microstructures, including echelle gratings and waveguides. The patterns are aligned to the same alignment marks used in the previous steps for defining the superconducting nanowires. Following development in xylenes, the grating and waveguide patterns are transferred to the Si$_3$N\textsubscript{4} film via carefully timed RIE in fluoroform (CHF\textsubscript{3}). The remaining resist is removed by hot NMP and gentle oxygen plasma. Afterwards, the AlO\textit{\textsubscript{x}} spacing layer and Al top ground layer are fabricated by another double-layer PMMA exposure in a fourth and final EBL step
and the following development, electron-beam evaporation and lift-off. The resulting devices are shown in Fig.$\,$2 and Supplementary Note 1.

\noindent \textbf{Optical grating and waveguide design.} 
To ensure high fabrication yield and verify the device principles, in this work we design relatively small gratings and short nanowires. The radius of Rowland circle for broadband device is \SI{400}{\micro\meter}, while the telecom-band device is $1.6\,\mathrm{mm}$. For the broadband design, we use small grating pitch (\SI{0.8}{\micro\meter}) and fundamental diffraction mode (\textit{m}=1) to minimize the order mixing. For the telecom-band design, we use a much larger grating pitch size (\SI{8}{\micro\meter}) and higher order mode (\textit{m}=6) to enhance the resolution and directivity. The width of the optical waveguide is \SI{1.2}{\micro\meter} to suppress TM modes and higher order TE modes for telecom wavelength around $1550\,\mathrm{nm}$. For efficient coupling with off-chip fibres, the input part of the waveguide is adiabatically tapered to \SI{4}{\micro\meter} to match the mode size of the lensed fibre.

\noindent \textbf{Detector and microwave circuit design. }The NbN nanowires on the Rowland circle are $60\,\mathrm{nm}$ wide, which is narrow enough to provide saturated detection efficiency of single photons over the whole spectrum as shown in Fig.$\,$3d. For some of the devices, two nanowires are connected in parallel to boost the signal-to-noise ratio (SNR) \cite{marsili_2011_SNAP_detector,cheng_2016_self_aligned_detector,cheng_2017_multiple_SNAP} while still maintaining the saturated internal detection efficiency (see Supplementary Note 1 and 5 for more details). The NbN nanowires are capped with 150$\,$nm-thick AlO\textit{\textsubscript{x}} and 150$\,$nm-thick Al to allow the nanowires to also function as microwave delay lines. As a result, a signal propagation velocity of as low as  0.73\% \textit{c} is experimentally obtained, where \textit{c} denotes the speed of light in vacuum. To preserve the fast-rising edges of the photon-excited detector pulses in readout circuits, we use Klopfenstein tapers for matching the impedance of the RF probe and cables to the nanowires by transforming the nanowire impedance to $50\,\Omega$. The total length of the tapers is $8.8\,\mathrm{mm}$, corresponding to 2.5 times effective wavelength of $1\,\mathrm{GHz}$ signal in the transmission line. 

\noindent \textbf{Device characterization.} The sample chip containing multiple spectrometer devices is cleaved after fabrication and then mounted on a 3-axis stack of stages (Attocube) inside a dilution refrigerator (BlueFors) and cooled down to $1.5\,\mathrm{K}$ temperature. We drive the stages to move the sample chip and make the electrode contact with a multi-channel RF probe. Another set of 3-axis stages (Attocube) are employed to align a single-mode tapered/lensed fibre (OZ Optics) to couple light into the devices from waveguide facet. This cryogenic active edge-coupling set-up guarantees efficient fibre-to-chip coupling over a broadband spectrum and also allows us to characterize multiple devices on one chip within a single cool-down experimental procedure. We use a broadband thermal light source (Thorlabs SLS201L) and a monochromator (Newport CS130) to generate 600-$1400\,\mathrm{nm}$ wavelength lines and employ tunable CW laser sources to generate other lines including 1480-$1640\,\mathrm{nm}$ (Santec TSL-710) and $1970\,\mathrm{nm}$ (Thorlabs TLK-L1900M). For timing jitter characterization, we use a 2.4$\,$ps-pulsed 1560$\,$nm laser (PolarOnyx). The arrival time difference (\textit{$\Delta t$}) of forward and backward photon-excited detector pulses are recorded by a $4\,\mathrm{GHz}$ and $40\,$GSample/s oscilloscope (Lecroy HDO9404). In order to minimize the extra timing jitter induced by the electronic noises from the readout circuits and thus maximize the spectral resolution, we employ two SiGe cryogenic low-noise amplifiers (Cosmic Microwave Technology CITLF1) operating at $4\,\mathrm{K}$ temperature to read out the photon-excited detector pulses (see Supplementary Note 5).

\section*{{\normalsize{}Data availability}}
\noindent The data that support the plots within this paper and other findings of this study are available from the corresponding author upon reasonable request. 

\begin{figure}[ht]
\centering
\includegraphics[width=0.95\linewidth]{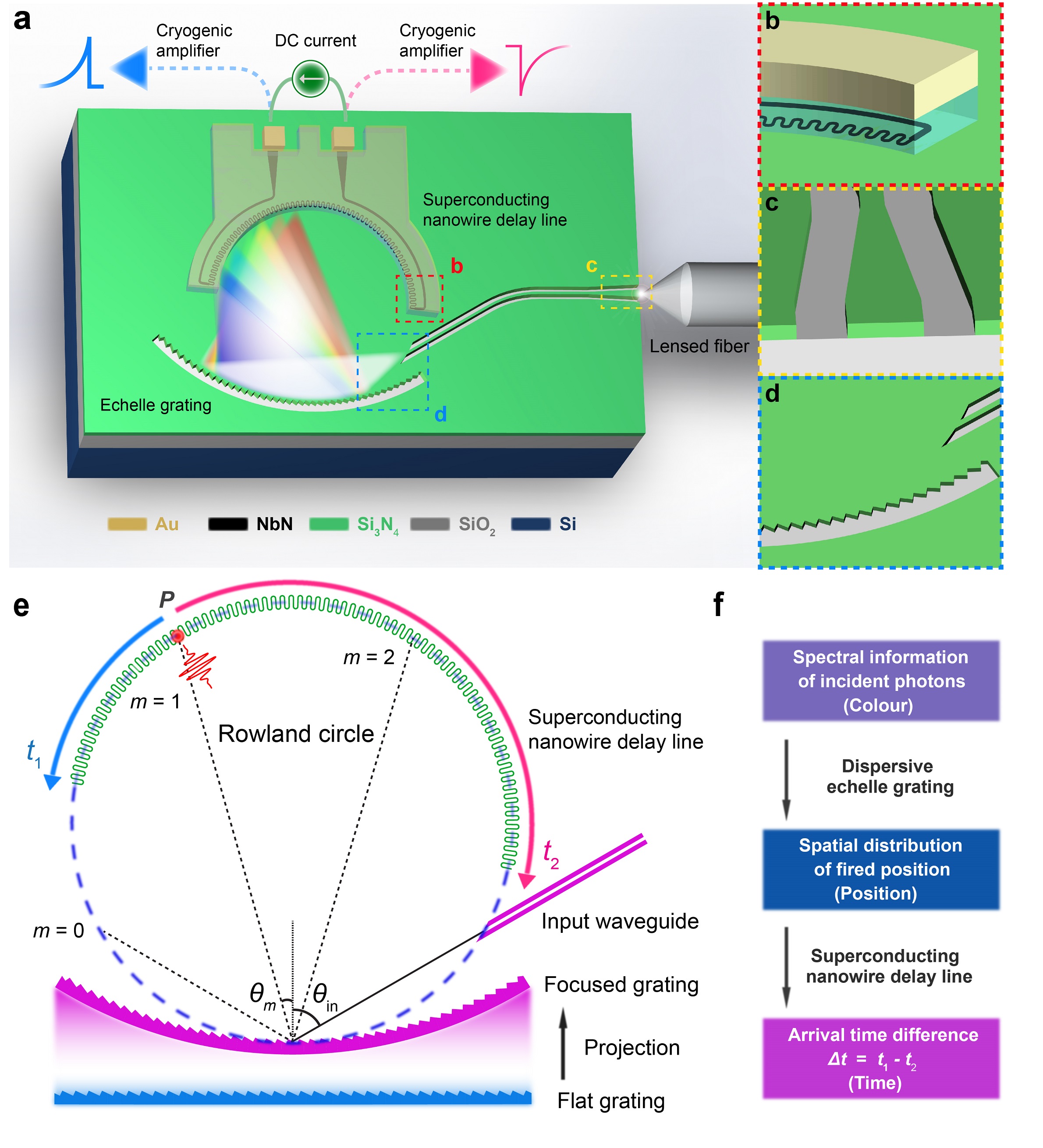}
\caption{\textbf{Device architecture and operation principle.} 
\textbf{a-d}, Three-dimensional sketch of the device. The on-chip focusing echelle grating operates as wavelength-discriminating microphotonic component while the superconducting nanowire functions simultaneously as a single-photon detector and a slow microwave delay line to continuously map the dispersed photons. The nanowire is capped with AlO\textit{\textsubscript{x}} as high-\textit{k} dielectric material and Al as top metal ground to form a slow microwave transmission line. 
\textbf{e}, Schematic illustration of the Rowland mounting. The input waveguide and the superconducting nanowire are mounted on the Rowland circle, which internally tangents with the focusing grating line. The focusing grating is projected  from a flat grating with the radius of the curvature equal to the diameter of the Rowland circle. 
\textbf{f}, Schematic representation of the signal transduction pathway and the device operation principle. 
}
\label{fig:stream}
\end{figure}


\begin{figure}[ht]
\centering
\includegraphics[width=\linewidth]{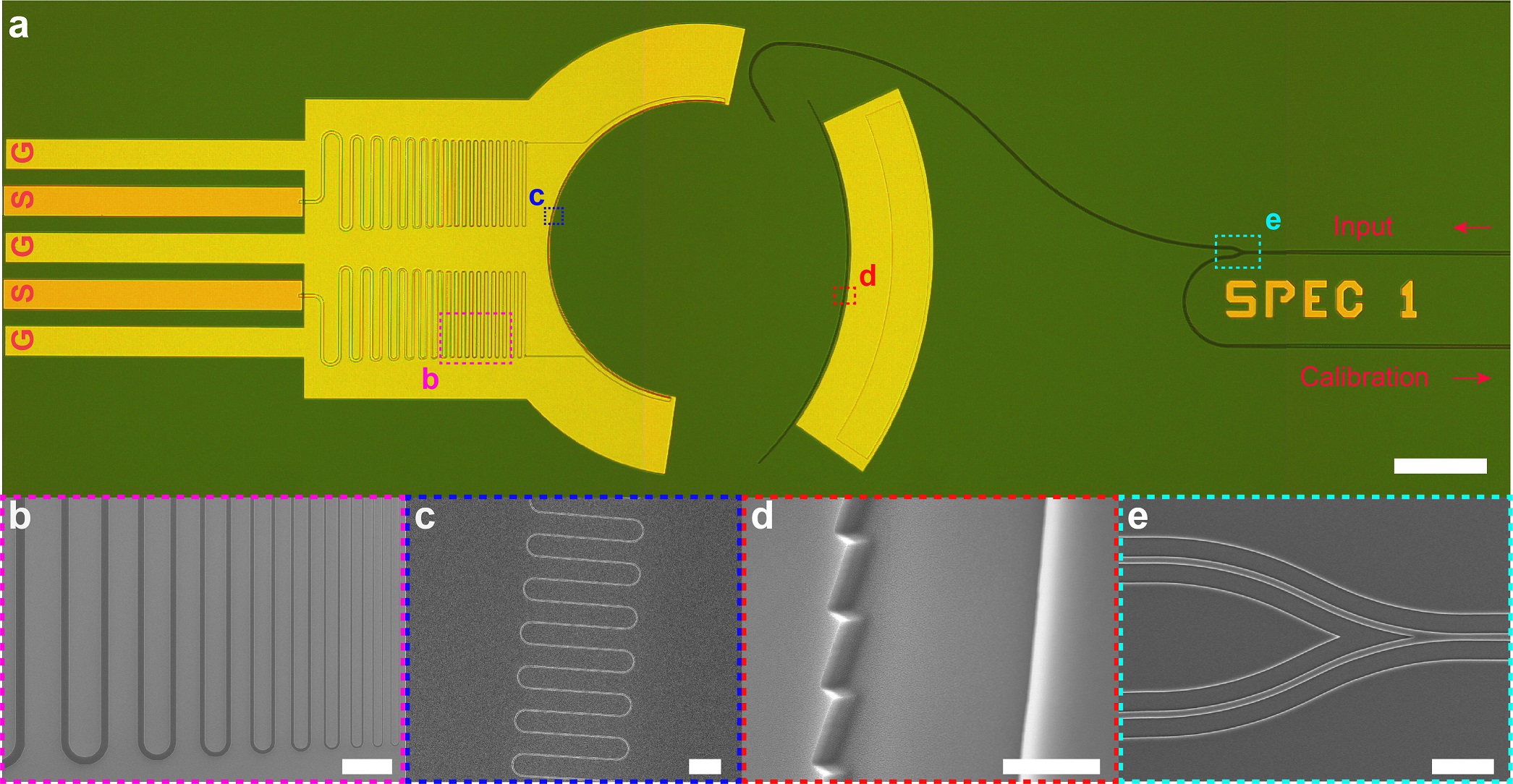}
\caption{\textbf{Broadband device images.} 
\textbf{a}, Overview optical micrograph image of the device. The input waveguide is first split into two waveguides, one to feed photons to the spectrometer, the other used for power calibration and monitoring the return power during the fibre-to-chip alignment. At the ends of the impedance tapers, the microstrip lines are converted to coplanar waveguides (CPWs) to match the modes of ground-signal-ground (GSG) RF probes. The yellow colour represents Al and AlO\textit{\textsubscript{x}} beneath, while the signal pads in orange colour are made of gold (Au). Scale bar, \SI{250}{\micro\meter}. 
\textbf{b}, Close-up SEM image of the impedance taper. Both ends of the nanowire detector are tapered from 60$\,$nm to microns width to preserve the fast-rising edges of photon-excited microwave pulses. Scale bar, \SI{10}{\micro\meter}.
\textbf{c}, Close-up SEM image of the meander nanowire detector. The nanowire is patterned from 8$\,$nm-thick NbN film. The width, pitch and depth of the nanowire is 60$\,$nm, 700$\,$nm and \SI{3}{\micro\meter}, respectively. Scale bar, \SI{1}{\micro\meter}. 
\textbf{d}, Angular SEM view of a section of the echelle grating. The pitch of the grating teeth is \SI{0.8}{\micro\meter} and the blaze angle is 20$\,$degree. Scale bar, \SI{1}{\micro\meter}. 
\textbf{e}, Expanded SEM view of the waveguide splitter. Scale bar, \SI{10}{\micro\meter}. The SEM images are taken prior to the deposition of AlO\textit{\textsubscript{x}} and Al for Device B (see Supplementary Note 1).}    
\label{fig:stream}
\end{figure}

\begin{figure}[ht]
\centering
\includegraphics[width=0.9\linewidth]{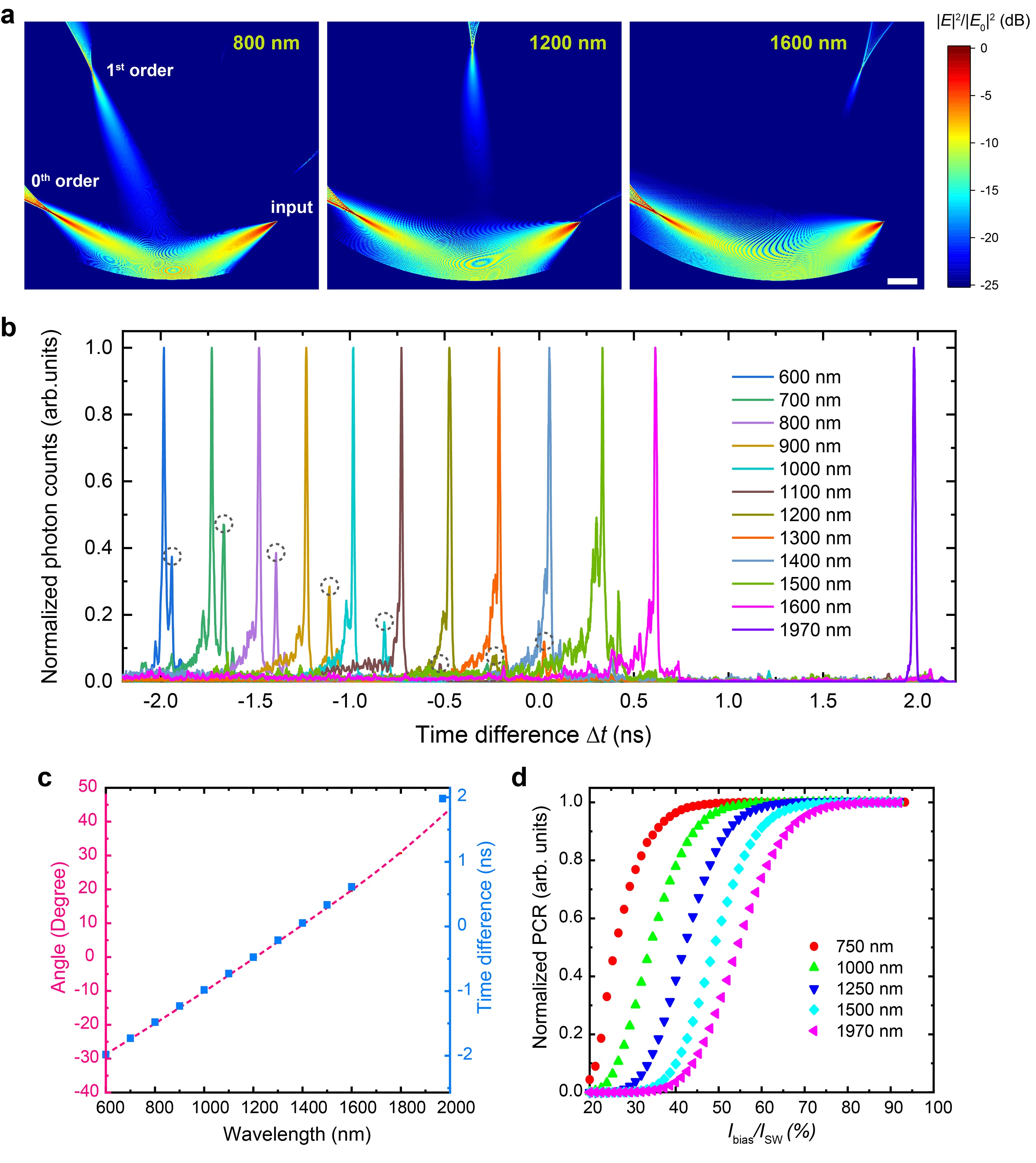}
\caption{\textbf{Broadband device results.} 
\textbf{a}, Electric field distribution of the device from 2.5-dimensional FDTD simulation results at different wavelengths. Scale bar, \SI{100}{\micro\meter}. 
\textbf{b}, Normalized histogram of photon counts versus time difference $\Delta t$ measured for different wavelength photons from 600$\,$nm to 1970$\,$nm. The main peaks are from the TE modes while the minor peaks marked by the dashed circles are from TM modes. The histogram is recorded with the nanowire detector biased at 80\% of its switching current $I_\mathrm{SW}$.
\textbf{c}, Comparison between experimentally measured $\Delta t$ and the diffraction angle extracted from the simulation results. 
\textbf{d}, Normalized photon counting rates (PCR) measured as a function of the bias current relative to the switching current of the device $I_\mathrm{bias}/I_\mathrm{SW}$ at wavelengths from 750$\,$nm to 1970$\,$nm. The complete saturation trend of the curves indicates a near-unity internal quantum efficiency of the nanowire detector over the whole spectrum.
The histogram results are taken from Device C with double-nanowire detector, while the efficiency curves are measured from Device B with single-nanowire structure for better comparison between different wavelengths (see Supplementary Note 1).
} 
\label{fig:stream}
\end{figure}

\begin{figure}[ht]
\centering
\includegraphics[width=\linewidth]{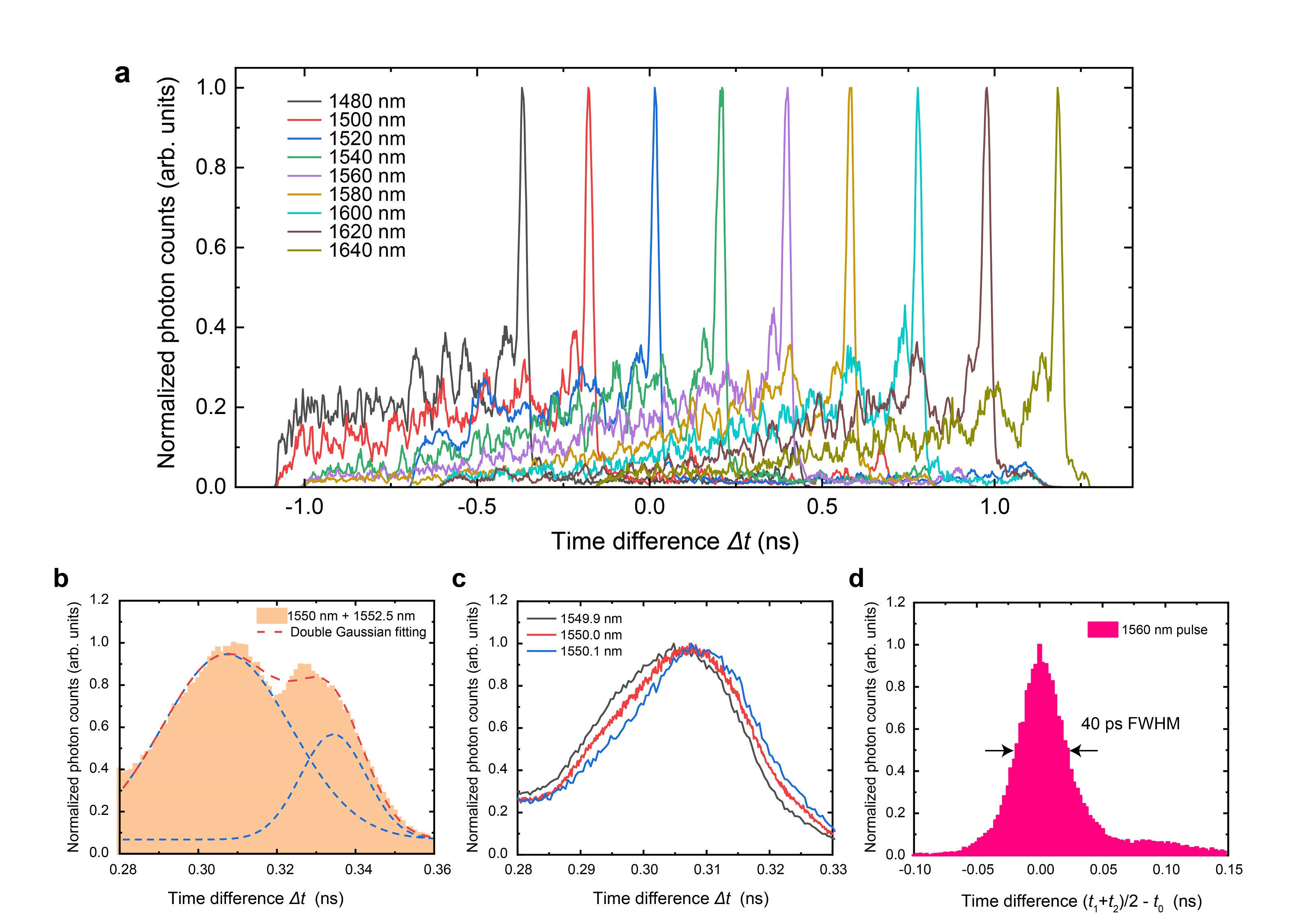}
\caption{\textbf{Telecom-band device results.} 
\textbf{a}, Normalized histogram of photon counts versus time difference $\Delta t$ measured for photons with different wavelength from 1480$\,$nm to 1640$\,$nm. 
\textbf{b}, Normalized histogram measured for a mixture of two coherent light sources with their wavelengths separated by 2.5$\,$nm. The red dashed line represents double-Gaussian fitting for the measured data, which is the sum of two Gaussian distribution displayed in blue dashed lines.
\textbf{c}, Normalized histogram measured for a single-colour source with slightly varied wavelength step by 0.1$\,$nm. 
\textbf{d}, Normalized histogram measured for a 1560$\,$nm pulsed source as a function of the arrival time difference between the averaged detector signal $(t_\mathrm{1}+t_\mathrm{2})/2$ and laser synchronization signal $t_\mathrm{0}$. All the histogram results are measured with the nanowire detector biased at 80\% of $I_\mathrm{SW}$.
All the results are taken from Device A (see Supplementary Note 1).
}
\label{fig:stream}
\end{figure}

\end{document}